%% file: main.tex
\DocumentMetadata{
  lang        = en-US,
  pdfversion  = 2.0,
  pdfstandard = UA-2,
  testphase   = phase-III,
}
\documentclass{article}
\makeatletter
\let\kernel@sect\@sect 
\makeatother
\usepackage{_stuffs/spconf,amsmath,graphicx,hyperref}

\makeatletter
\let\@sect\kernel@sect
\renewcommand\@seccntformat[1]{\csname the#1\endcsname.\hskip 0.6em}
\renewcommand\section{\@startsection{section}{1}{\z@}%
  {-3.5ex \@plus -1ex \@minus -.2ex}{2.3ex \@plus.2ex}{\centering\bfseries}}
\renewcommand\subsection{\@startsection{subsection}{2}{\z@}%
  {-3.25ex\@plus -1ex \@minus -.2ex}{1.5ex \@plus .2ex}{\bfseries}}
\renewcommand\subsubsection{\@startsection{subsubsection}{3}{\z@}%
  {-3.25ex\@plus -1ex \@minus -.2ex}{1.5ex \@plus .2ex}{\itshape}}
\NewCommandCopy\spconf@section\section
\RenewDocumentCommand\section{s o m}{%
  \IfBooleanTF{#1}{\spconf@section*{\MakeUppercase{#3}}}%
    {\IfNoValueTF{#2}{\spconf@section[#3]{\MakeUppercase{#3}}}%
                     {\spconf@section[#2]{\MakeUppercase{#3}}}}}
\def\@maketitle{\newpage
 \null
 \vskip 2em \begin{center}
 {\large \bf \tagstructbegin{tag=Title}\@title \par\tagstructend} \vskip 1.5em {\large \lineskip .5em
 \@name \vskip\baselineskip \@address \par} \end{center}
 \par
 \vskip 1.5em}

\ExplSyntaxOn
\socket_if_exist:nT {caption/label}
  {
    \NewSocketPlug{caption/label}{spconf}{#1.~}
    \AssignSocketPlug{caption/label}{spconf}
  }
\ExplSyntaxOff
\makeatother
\usepackage{textcomp}
\usepackage{booktabs}
\usepackage{pifont}
\usepackage{xspace}
\usepackage{xcolor}
\usepackage{enumitem}
\usepackage{algorithm}
\usepackage{algpseudocode}
\usepackage{microtype}
\usepackage{array}
\usepackage{tabularx}
\usepackage{colortbl, xcolor}
\usepackage{afterpage}
\usepackage{multirow}

\ExplSyntaxOn
\NewDocumentCommand\tabletagsetup{m}
  {\keys_if_exist:nnT{__tag/setup}{table/header-rows}{\tagpdfsetup{#1}}}
\ExplSyntaxOff

\hypersetup{
  pdftitle={Auditory Illusion Benchmark for Large Audio Language Models},
  pdfauthor={Hayoon Kim, Eunice Hong, Kyogu Lee},
  pdfsubject={Benchmark of auditory illusion susceptibility in large audio language models},
  pdfkeywords={Illusion, Auditory Illusion, Large Audio Language Models, Benchmark, Dataset},
  pdfdisplaydoctitle=true,
}
\ifdefined\pdfglyphtounicode \pdfglyphtounicode{a19}{2713}\fi

\renewcommand\paragraph[1]{\noindent\textbf{#1.}}

\title{Auditory Illusion Benchmark for Large Audio Language Models}
\name{
    Hayoon Kim\textsuperscript{$\flat\natural$}, 
    Eunice Hong\textsuperscript{$\flat\natural$}, 
    Kyogu Lee\textsuperscript{$\flat\natural\sharp\dagger$}
}

\address{
    \textsuperscript{$\flat$}Music and Audio Research Group, Seoul National University\\
    \textsuperscript{$\natural$}Department of Intelligence and Information, Seoul National University \\    
    \textsuperscript{$\sharp$}AIIS, Seoul National University \quad
    \textsuperscript{$\dagger$}IPAI, Seoul National University \\
    \texttt{\{hyway, euniunie, kglee\}@snu.ac.kr}
}

\begin{document}
\ninept

\maketitle

\begin{abstract}
Perceptual illusions have long served as crucial probes into human cognition, revealing biases and limitations of perception. In the auditory domain, such illusions provide a unique lens for testing whether Large Audio Language Models (LALMs) replicate human perceptual tendencies. Despite their importance, most benchmarks focus on visual illusions or general audio tasks, leaving auditory illusions underexplored. To this end, we present \texttt{AIB}, the first auditory illusion benchmark for LALMs, covering ten representative illusions across music, sound, and speech, each annotated for the presence of knowledge-based priors. Our methodology pairs model evaluation with controlled human listening studies, enabling direct comparison of responses. Results show systematic differences: while most LALMs remain signal-faithful on low-level acoustic illusions, several exhibit more human-like responses when linguistic or musical priors are involved, although no model matches the human perceptual profile. These findings highlight the current limitations of LALMs as cognitive models. By establishing auditory illusions as a rigorous testbed, our work offers a new perspective for probing neural black-box models and advancing understanding of auditory cognition. \texttt{AIB} is publicly available at \url{https://github.com/gillosae/aib}.
\end{abstract}

\begin{keywords}
Illusion, Auditory Illusion, Large Audio Language Models, Benchmark, Dataset
\end{keywords}

\input{section/1-intro}
\input{section/2-related-works}
\input{section/3-benchmark-dataset}
\input{section/4-experiment}
\input{section/5-result}
\input{section/6-discussion}

\input{section/7-conclusion}

\section{Acknowledgements}
This work was partly supported by the National Research Foundation of Korea (NRF) grant funded by the Korea government (MSIT) [No. RS-2024-00461617, 50\%], [No. RS-2025-24683892, 45\%], and Institute of Information \& communications Technology Planning \& Evaluation (IITP) grant funded by the Korea government (MSIT) [NO.RS-2021-II211343, 5\%]. The authors would also like to thank Kihong Kim for his helpful discussions and support.



\bibliographystyle{_stuffs/IEEEbib}
\bibliography{refs}

\end{document}

%% file: section/1-intro.tex
\section{Introduction}
\label{sec:intro}
Perceptual illusions offer cognitive science insight into the mechanisms and biases underlying human perception \cite{Eagleman2001, choi2025effects, hong2025concurrent}. In the auditory domain, classic illusions such as the Shepard tone \cite{shepard1964circularity} or the missing fundamental \cite{chialvo2003we} illustrate how the brain integrates acoustic signals and prior knowledge to form coherent percepts that diverge from the physical stimulus \cite{warren1971speech}, revealing both the constructive nature of auditory perception and its systematic fallibility. 

Recent advances have produced Large Audio Language Models (LALMs) with impressive capabilities in automatic speech recognition, audio captioning, and music understanding. 
Although these models show limited success in tasks with clear ground truth \cite{zang2025really}, little is known about whether they also internalize human-like perceptual biases, including susceptibility to illusions. 

\input{figures/example}

Existing audio and multimodal benchmarks focus primarily on tasks with objective, signal-level ground truth, such as transcription accuracy, 
classification, or retrieval \cite{Gemmeke2017AudioSet}.
In contrast, illusions pose a fundamentally different question: whether systems replicate the same subjective misperceptions observed in humans, a property already explored for computer vision models but not yet systematically investigated in the auditory domain.

To address this gap, we introduce the first auditory illusion benchmark (\texttt{AIB}) for LALMs. \texttt{AIB} spans ten representative illusions from prior literature, categorized along two axes: underlying mechanism (physics-based vs. physics+knowledge-based) and perceptual content (music, sound, speech). Figure~\ref{fig:teaser} shows example prompts. We pair model evaluations with controlled human listening studies, enabling direct comparison between models and humans.

Our results show systematic divergences between humans and LALMs. For low-level acoustic illusions, humans consistently exhibited strong susceptibility, while models ranged from partial alignment to strict adherence to the physical signal. In contrast, illusions that additionally rely on linguistic or musical priors elicited more human-like responses in several models, although no model matched the human perceptual profile. Notably, some models shifted from signal-faithful behavior on physics-based illusions to greater illusion susceptibility on knowledge-driven ones, suggesting that this alignment may arise from higher-level learned priors rather than shared low-level auditory processing.

First, we introduce a benchmark that establishes auditory illusions as a new dimension of LALM evaluation, together with the open-source implementations for systematically generating the benchmark illusions. Second, we provide human baseline data through controlled perceptual studies. Third, we present empirical analyses showing that LALM illusion susceptibility diverges systematically from humans, highlighting both progress in perceptual modeling and persistent gaps in human-likeness.
By situating auditory illusions as a rigorous testbed, our work extends evaluation beyond recognition accuracy and towards perceptual alignment, contributing to both machine learning and cognitive science perspectives on auditory perception.

%% file: figures/example.tex
\begin{figure}
    \centering
    \includegraphics[width=\linewidth, alt={Three example prompts from the benchmark ("Were the perceived notes of the two audios the same?", "Did you hear an aftertone when the noise stopped?", "Did you hear a low D3 tone in addition to G5 and A sharp 5?"). An audio stimulus is played to both a human listener and a large audio language model, and their yes/no answers are compared.}]{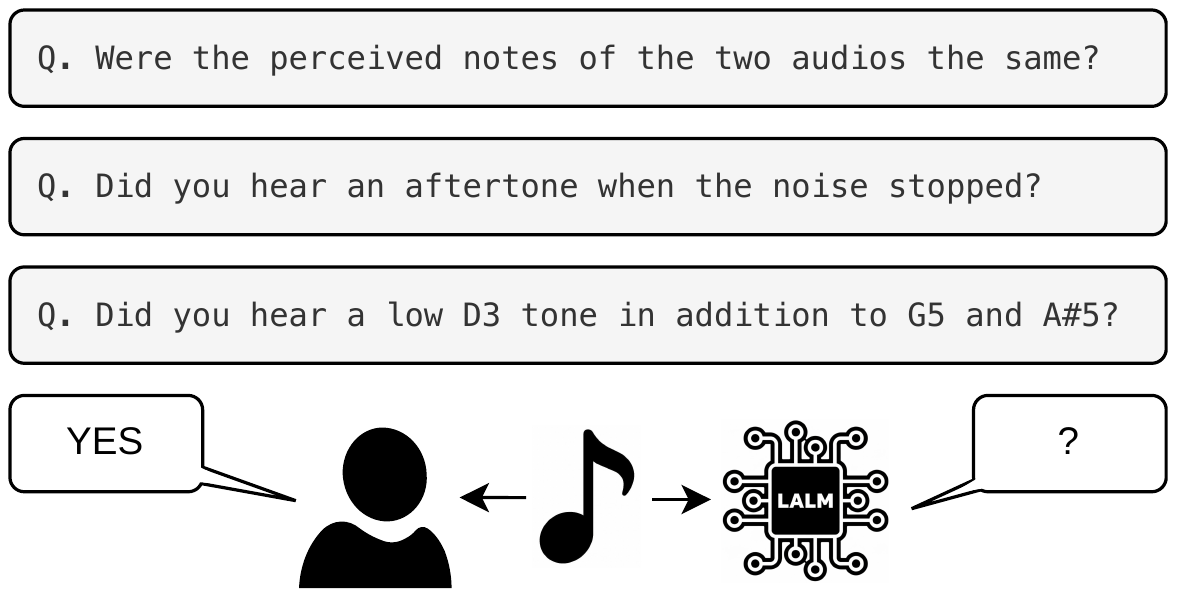}
    \vspace{-6mm}
    \caption{
    Representative samples from the \texttt{AIB} benchmark, featuring both illusion-inducing (e.g., Missing fundamental, Zwicker tone, and Tartini tone) and control stimuli. The benchmark queries human participants and AI models on their perceptual experiences to evaluate how closely AI responses align with human-like auditory perception.
    }
    \label{fig:teaser}
    \vspace{-4mm}
\end{figure}

%

%% file: section/2-related-works.tex
\section{Related Work}
\label{sec:related_works}

\paragraph{Auditory Illusions}
Auditory illusions reveal systematic mismatches between acoustic input and perception, providing powerful probes into auditory mechanisms. Early accounts, such as Tartini's third tone, were later explained by Helmholtz as non-linear responses of the auditory system \cite{carson2007musical}. Modern systematic research was pioneered by Deutsch, who discovered the octave \cite{deutsch1981octave}, scale \cite{deutsch1975musical}, and tritone paradox \cite{deutsch1991tritone}, showing variation across brain organization, handedness, and linguistic background. Other canonical examples include the missing fundamental, Shepard tone, Risset rhythm \cite{risset1978paradoxes}, Zwicker tone \cite{franosch2003zwicker}, and temporal gap illusion \cite{formby1991detection}, often interpreted via Gestalt principles of closure \cite{scharine2009auditory}. 
Speech-related illusions such as phonemic restoration, speech-to-song \cite{deutsch2011illusory, joung2025exploring}, phantom words, and high-profile cases like the Yanny/Laurel debate \cite{leemann2022factors} further highlight how perception imposes linguistic structure on ambiguous inputs. These findings establish illusions as rigorous tools for studying auditory cognition, though certain binaural or human-specific illusions remain difficult to model computationally. Beyond their role in basic psychoacoustics, auditory illusions have also been used in neuroscience to study neural adaptation \cite{husain2005investigating}, in HCI to probe auditory attention and perceptual learning \cite{Eagleman2001}, and in cross-linguistic settings to reveal how cultural and linguistic priors shape auditory interpretation \cite{deutsch1991tritone}. This broad applicability underscores their value as diagnostic tools not only for human perception but also for evaluating artificial systems.

\vspace{2mm}
\paragraph{Large Audio Language Models}
LALMs extend transformer-based architectures into the auditory domain by combining multimodal encoders with large-scale training \cite{radford2022robust}.
AudioPaLM \cite{rubenstein2023audiopalm} unified recognition and translation by transferring linguistic and paralinguistic knowledge, while Qwen-Audio \cite{chu2023qwen} supported multitask understanding across speech, music, and environmental audio. Audio Flamingo \cite{kong2024audio} further extended these capabilities to in-context dialogue about audio. More recent work has emphasized reasoning: GAMA-IT \cite{ghosh2024gama} used synthetic datasets for instruction tuning to reduce hallucinations, and Audio-Reasoner \cite{xie2025audio} introduced CoTA \cite{ma2025audiocot} to enhance inference in audio QA. While these systems demonstrate strong performance across diverse auditory tasks, their evaluation remains primarily recognition-oriented, leaving open whether they capture human-like perceptual fallibilities. These systems differ not only in training objectives but also in their inductive biases: dual-encoder designs emphasize modality alignment, while unified encoder-decoders encourage transfer across tasks. Reasoning-tuned models such as GAMA or Audio-Reasoner suggest that chain-of-thought methods can reduce hallucination, yet none of these approaches directly evaluate whether models internalize human-like perceptual biases such as illusions. Addressing this gap is critical if LALMs are to serve as cognitive models of auditory processing.

\vspace{2mm}
\paragraph{Auditory or Illusion Benchmarks}
Benchmarking in audio has evolved from recognition to reasoning and instruction following. SUPERB \cite{yang2021superb} 
evaluates a wide range of speech and audio tasks, while AudioBench \cite{wang2024audiobench} unifies evaluation across 26 datasets with both objective metrics and LLM-based judging. Recent efforts such as MMAU \cite{sakshi2024mmau}, AIR-Bench \cite{yang2024air}, MMAR \cite{ma2025mmar} 
emphasize inference and multi-step reasoning across speech, music, and environmental audio. CoTA \cite{xie2025audio} provides reasoning-oriented captions and chain-of-thought annotations to advance emergent reasoning.  
Despite their breadth, these benchmarks remain anchored in tasks with objective ground truth labels, such as classification or transcription. They cannot assess whether models replicate systematic human misperceptions. In contrast, illusion-focused benchmarks in vision \cite{zhang2025illusionbench, guan2024hallusionbench, zhang2023grounding} explicitly probe subjective misalignment between models and humans.

In vision-language research, illusion-focused benchmarks have directly tested perceptual biases. IllusionBench \cite{zhang2025illusionbench} demonstrated that larger VLMs align more closely with human susceptibility to classic illusions, while HallusionBench \cite{guan2024hallusionbench} revealed persistent mismatches where models hallucinate or miss illusions. These works highlight illusions as probes of perceptual alignment. However, no systematic auditory benchmark currently examines whether LALMs perceive auditory illusions like humans. Our work addresses this gap by introducing the first auditory illusion benchmark and comparing model predictions with human judgments.

%% file: section/3-benchmark-dataset.tex
\section{Benchmark Dataset}
\label{sec:benchmark_dataset}

\input{tables/illusion_category}

\paragraph{Overview}
The \texttt{AIB} benchmark dataset comprises ten representative auditory illusions, including 10,385 illusion stimuli and 4,444 control stimuli, as detailed in Table~\ref{tab:illusion_category}. Each illusion belongs to one of three domains---music, speech, or general sound---and is further categorized by underlying mechanism (physics-based vs.\ physics+knowledge-based) as detailed in Section~\ref{subsec:data_curation}. For every illusion, both stimulus and matched control sets were collected to ensure controlled conditions, enabling systematic comparison of human judgments and model predictions.

\vspace{2mm}
\paragraph{Data Curation and Illusion Classification}
\label{subsec:data_curation}
We compiled a set of academically recognized auditory illusions for this study, excluding those that fundamentally depend on binaural physiology (e.g., the octave illusion) or on mechanisms beyond the representational capacity of current computational models, such as spatial-hearing–specific effects \cite{deutsch1981octave}. This ensured that all selected illusions could be robustly presented, measured, and compared across both human listeners and available LALMs.

Although auditory illusions have been extensively investigated \cite{deutsch2011illusory}, systematic classification has been more thoroughly developed in the visual domain. In vision science, Gestalt principles have provided influential taxonomies of illusion types \cite{coren1980principles}. However, for our purposes, Gregory’s causal framework—distinguishing illusions by their origins—offers a more direct mapping to auditory phenomena \cite{gregory1997knowledge}.
Gregory categorized illusions into two groups: those arising from physical or physiological factors (e.g., optical distortions, sensory signal disturbances) and those driven by knowledge, rules, or prior experience. The former are stimulus-driven and bottom-up, whereas the latter emerge from top-down processes that impose expectations or contextual knowledge on ambiguous input \cite{gregory1997visual}.

Adapting this framework to the auditory domain, we define two categories: (i) \textit{Physics-based auditory illusions}, where percepts arise from acoustic properties or basic auditory physiology, including frequency interactions, masking, and temporal adaptation; and (ii) \textit{Physics+Knowledge-based auditory illusions}, where percepts are also shaped by prior knowledge or expectations, such that top-down mechanisms impose linguistic, musical, or contextual structure on incomplete input. Each illusion in our benchmark was systematically assigned to one of these categories based on its dominant causal mechanism, following Gregory’s distinction between bottom-up and top-down processes in perception.

We emphasize that this framework reflects the dominant explanatory mechanism rather than a strict dichotomy. In ambiguous cases, classification is based on the dominant causal contribution rather than exclusivity. This classification strategy ensures that illusions can be robustly benchmarked across both human listeners and artificial systems.

%% file: tables/illusion_category.tex
\begin{table*}[t]
  \centering
  \caption{Statistical summary of the \texttt{AIB} dataset. For each illusion we list its domain, its dominant mechanism (P: Physics; P+K: Physics+Knowledge), and the number of illusion and control stimuli in the full benchmark and in the mini subset.}
  \label{tab:illusion_category}
  {\fontsize{8.5}{10.5}\selectfont
  \setlength{\tabcolsep}{5pt}
  \renewcommand{\arraystretch}{1.08}
  \tabletagsetup{table/header-rows={1,2}, table/header-columns={1,2}}
  \begin{tabular}{@{}llwc{1.6em}wc{2.2em}rrrrrr@{}}
    \toprule
    \tabletagsetup{table/multirow=2}\multirow{2}{*}[-2.5pt]{\textbf{Domain}} & \tabletagsetup{table/multirow=2}\multirow{2}{*}[-2.5pt]{\textbf{Illusion}}
      & \multicolumn{2}{c}{\textbf{Mechanism}}
      & \multicolumn{3}{c}{\textbf{Full set}}
      & \multicolumn{3}{c@{}}{\textbf{Mini set}} \\
    \cmidrule(lr){3-4} \cmidrule(lr){5-7} \cmidrule(l){8-10}
      & & P & P+K & Illusion & Control & Total & Illusion & Control & Total \\
    \midrule
    \tabletagsetup{table/multirow=4}\multirow{4}{*}{Music}
      & Missing fundamental \cite{chialvo2003we}     & \ding{51} &           & 1,080 &   178 & 1,258 & 108 &  88 & 196 \\
      & Risset rhythm \cite{risset1978paradoxes}     &           & \ding{51} &   960 &   144 & 1,104 &  96 &  23 & 119 \\
      & Shepard tone \cite{shepard1964circularity}   &           & \ding{51} & 1,296 &   518 & 1,814 & 127 &  56 & 183 \\
      & Tartini tone \cite{meyer1957subjective}      & \ding{51} &           &   720 &   179 &   899 &  72 &  10 &  82 \\
    \midrule
    \tabletagsetup{table/multirow=4}\multirow{4}{*}{Sound}
      & Temporal gap \cite{formby1991detection}      &           & \ding{51} & 1,350 &   540 & 1,890 & 135 &  48 & 183 \\
      & Zwicker tone \cite{franosch2003zwicker}      & \ding{51} &           &   432 &    61 &   493 &  43 &  43 &  86 \\
      & Tempo change illusion \cite{boltz2011illusory} &         & \ding{51} &   896 &   537 & 1,433 &  89 &  66 & 155 \\
      & Continuity illusion \cite{husain2005investigating} &     & \ding{51} &   432 &   172 &   604 &  36 &  16 &  52 \\
    \midrule
    \tabletagsetup{table/multirow=2}\multirow{2}{*}{Speech}
      & Speech-to-song \cite{deutsch2011illusory}    &           & \ding{51} & 2,304 & 1,766 & 4,070 & 234 & 150 & 384 \\
      & Phonemic restoration \cite{warren1971speech} &           & \ding{51} &   915 &   349 & 1,264 &  92 &  34 & 126 \\
    \midrule
    \textbf{Total} & \textbf{10 illusions} & \textbf{3} & \textbf{7}
      & \textbf{10,385} & \textbf{4,444} & \textbf{14,829} & \textbf{1,032} & \textbf{534} & \textbf{1,566} \\
    \bottomrule
  \end{tabular}}
\end{table*}

%% file: section/4-experiment.tex
\vspace{-.5mm}
\section{Experiments}
\vspace{-.5mm}
\label{sec:experiment}

\paragraph{Evaluation Setup}
We evaluated a set of LALMs alongside proprietary baselines, which have demonstrated strong performance on multimodal audio benchmarks. Models were presented with the same stimuli as human listeners. To adapt illusions into machine-interpretable tasks, we reformulated each perceptual judgment as a multiple-choice question, limited to two formats: (i) \textit{binary choices}, such as yes/no or same/different judgments, and (ii) \textit{ternary choices}, such as ascending/descending/no change for pitch-related illusions. This setup ensures direct comparability with human responses and follows the design principles of prior illusion benchmarks \cite{zhang2025illusionbench, guan2024hallusionbench, zhang2023grounding}, adapted here to the auditory domain. Model outputs were parsed into the answer set; unparseable responses were excluded.

\vspace{2mm}
\paragraph{Metrics}
Model responses were compared against human response distributions. We report:
    Human-Likeness Accuracy (HLA): proportion of model outputs matching the dominant human response.
    Reality Alignment (RA): proportion of outputs aligned with the physical ground truth (illusion-free).
    Illusion Susceptibility Index (ISI): difference between HLA and RA, quantifying the extent to which models share human-like misperceptions. We frame ISI as a descriptive measure of human-likeness rather than a monolithic performance metric; high or low ISI values are task-dependent and should be interpreted in the context of the specific application goals.

\vspace{2mm}

\paragraph{Human Listening Study}
To establish perceptual baselines, we recruited 20 participants possessing absolute pitch for controlled human experiments. Absolute-pitch listeners were chosen because several illusions (e.g., Shepard and Tartini tones) require reliable note-level judgments, allowing us to separate perceptual susceptibility from pitch-labeling error. Participants completed a randomized trial sequence that included illusion and control stimuli, presented without feedback so that responses reflected immediate percepts rather than learned strategies. The experiment was conducted via a web-based interface, and participants were instructed to use headphones in a quiet environment to ensure signal clarity. For each trial, they reported categorical judgments (e.g. `up' vs.\ `down' in Shepard tones, `complete' vs.\ `incomplete' in phonemic restoration). To ensure data reliability, we aggregated responses using majority voting to construct robust ground-truth distributions for each illusion.

%% file: section/5-result.tex
\vspace{-.5mm}
\section{Results}
\vspace{-.5mm}
\label{sec:result}
\input{tables/benchmark_results_full}
Table~\ref{tab:benchmark_results} summarizes our evaluation. Since the human reference defines HLA and ISI as 1.0 by construction, the question is how far each model departs from it, and whether departures reflect a genuine percept or a response bias, which the matched controls diagnose. No model approaches the human profile: the best average ISI (Audio Flamingo 3, 0.455) is less than half the human reference, and control accuracy is modest throughout, with several models below chance. ISI must therefore be read jointly with control accuracy: a degenerate strategy that always gives the illusion-consistent answer attains an average ISI of 0.799 while failing Physics controls almost entirely. This instability indicates that robust auditory understanding in LALMs remains an open challenge, but also positions the benchmark as a forward-looking diagnostic.

On physics-based illusions, only MuLLaMa and Audio Flamingo 3 share the human-like percept, and both do so with below-chance control accuracy, so part of their susceptibility is answer bias rather than perception. All remaining models side with the physical signal, and the most signal-faithful models are also those with the highest control accuracy: they act as detectors of what is physically present (e.g., the absent fundamental) rather than of what a listener hears.

Illusions that additionally rely on top-down priors elicit the most human-like behavior. Audio Flamingo 3 attains the highest ISI (0.505) with above-chance controls, and Gemini 3.1 Pro the highest HLA (0.707). Notably, Gemini 3.1 Pro, Kimi-Audio-Instruct, and Fun-Audio-Chat reverse sign between domains: strongly signal-faithful on acoustic illusions yet human-like once linguistic or musical expectations are involved, suggesting alignment driven by language-level priors rather than shared low-level auditory processing. Qwen2-Audio-Instruct is the most consistently literal model, pairing negative ISI in both domains with the highest control accuracy, whereas GLM-4-Voice is simply unstable, with negative ISI and far below-chance controls.

Re-evaluating three checkpoints with refined prompts (last block of Table~\ref{tab:benchmark_results}) shows that wording alone can shift ISI by as much as 0.6, more than most between-model differences: Qwen2-Audio-Instruct collapses to a purely physical reading with perfect Physics control accuracy. Susceptibility of instruction-tuned models is thus a property of the model-prompt pair rather than of the checkpoint alone.

Across domains, three regimes emerge: susceptible models (Audio Flamingo 3, MuLLaMa) with positive ISI but weak controls; literal models (Qwen2-Audio-Instruct, Voxtral-Mini, Qwen-Audio-Chat) with negative ISI and above-chance controls; and domain-dependent models (Gemini 3.1 Pro, Kimi-Audio-Instruct) that are literal on acoustic illusions but susceptible on knowledge-driven ones. No model combines high susceptibility with reliable controls, which is precisely the human profile. Since same-scale models span the entire ISI range, progress likely requires not scale alone but training that integrates perceptual priors with reliable signal-level judgment.

%% file: tables/benchmark_results_full.tex
\begin{table*}[t]
  \centering
  \caption{Performance of all evaluated LALMs on auditory illusions, categorized by Physics and Physics+Knowledge domains. All retained illusion stimuli were validated as eliciting the intended percept in the human listening study, so the human response becomes the illusion label. $\dagger$ marks models whose free-text replies could not be mapped onto the answer set for more than 20\% of trials; their scores understate failure rather than measuring perception. The highest ISI and control accuracy among the evaluated models in each column are shown in bold.}
  \label{tab:benchmark_results}
  {
  \fontsize{9}{11}\selectfont
  \setlength{\tabcolsep}{3pt}
  \renewcommand{\arraystretch}{1.05}
  \tabletagsetup{table/header-rows={1,2,3}, table/header-columns={1}}
  \scalebox{0.93}{
  \begin{tabular}{>{\raggedright\arraybackslash}p{5.2cm}cccrcccrcccrc}
    \toprule
    \tabletagsetup{table/multirow=3}\multirow{3}{*}[-4.5pt]{\textbf{Models}} & \tabletagsetup{table/multirow=3}\multirow{3}{*}[-4.5pt]{\textbf{Size}}
    & \multicolumn{4}{c}{\textbf{Physics}}
    & \multicolumn{4}{c}{\textbf{Physics+Knowledge}}
    & \multicolumn{4}{c}{\textbf{Average}} \\
    \cmidrule(lr){3-6} \cmidrule(lr){7-10} \cmidrule(lr){11-14}
    & & \multicolumn{3}{c}{Illusion} & \tabletagsetup{table/multirow=2}\multirow{2}{*}[-2.25pt]{Control}
    & \multicolumn{3}{c}{Illusion} & \tabletagsetup{table/multirow=2}\multirow{2}{*}[-2.25pt]{Control}
    & \multicolumn{3}{c}{Illusion} & \tabletagsetup{table/multirow=2}\multirow{2}{*}[-2.25pt]{Control} \\
    \cmidrule(lr){3-5} \cmidrule(lr){7-9} \cmidrule(lr){11-13}
    & & HLA & RA & \multicolumn{1}{c}{ISI} & & HLA & RA & \multicolumn{1}{c}{ISI} & & HLA & RA & \multicolumn{1}{c}{ISI} & \\
    \midrule
    \multicolumn{14}{l}{\textit{Baselines and human reference}} \\
    Random Guess & -- & 0.500 & 0.500 & 0.000 & 0.500 & 0.436 & 0.436 & 0.000 & 0.450 & 0.449 & 0.449 & 0.000 & 0.455 \\
    Most Frequent Choice & -- & 1.000 & 0.000 & 1.000 & 0.110 & 0.744 & 0.000 & 0.744 & 0.607 & 0.799 & 0.000 & 0.799 & 0.560 \\
    Human & -- & 1.000 & 0.000 & 1.000 & 1.000 & 1.000 & 0.000 & 1.000 & 1.000 & 1.000 & 0.000 & 1.000 & 1.000 \\
    \midrule
    \multicolumn{14}{l}{\textit{Large Audio Language Models (LALMs)}} \\
    Pengi \cite{deshmukh2023pengi}$^{\dagger}$ & 323M & 0.000 & 0.014 & $-$0.014 & 0.014 & 0.278 & 0.177 & 0.101 & 0.153 & 0.218 & 0.142 & 0.076 & 0.140 \\
    Voxtral-Mini \cite{liu2025voxtral} & 3B & 0.463 & 0.478 & $-$0.015 & 0.526 & 0.357 & 0.509 & $-$0.152 & 0.602 & 0.380 & 0.503 & $-$0.123 & 0.595 \\
    MuLLaMa \cite{liu2024music} & 7B & 0.659 & 0.341 & \textbf{0.319} & 0.368 & 0.481 & 0.274 & 0.207 & 0.432 & 0.519 & 0.288 & 0.231 & 0.426 \\
    Kimi-Audio-Instruct \cite{ding2025kimi} & 7B & 0.124 & 0.876 & $-$0.752 & 0.782 & 0.540 & 0.279 & 0.262 & 0.621 & 0.451 & 0.407 & 0.044 & 0.636 \\
    Audio Flamingo 3 \cite{kong2024audio} & 8B & 0.584 & 0.313 & 0.271 & 0.404 & 0.598 & 0.093 & \textbf{0.505} & 0.592 & 0.595 & 0.140 & \textbf{0.455} & 0.574 \\
    Fun-Audio-Chat \cite{team2025fun} & 8B & 0.008 & 0.676 & $-$0.668 & 0.553 & 0.450 & 0.410 & 0.039 & 0.589 & 0.355 & 0.467 & $-$0.112 & 0.585 \\
    Qwen-Audio-Chat \cite{chu2023qwen} & 8.4B & 0.427 & 0.573 & $-$0.146 & 0.548 & 0.345 & 0.381 & $-$0.037 & 0.512 & 0.362 & 0.423 & $-$0.060 & 0.516 \\
    Qwen2-Audio-Instruct \cite{chu2024qwen2} & 8.4B & 0.228 & 0.769 & $-$0.541 & 0.727 & 0.335 & 0.565 & $-$0.230 & \textbf{0.642} & 0.312 & 0.609 & $-$0.297 & \textbf{0.650} \\
    GLM-4-Voice \cite{zeng2024glm} & 9B & 0.321 & 0.485 & $-$0.164 & 0.488 & 0.399 & 0.582 & $-$0.184 & 0.227 & 0.382 & 0.561 & $-$0.179 & 0.251 \\
    \midrule
    \multicolumn{14}{l}{\textit{Large Language Models (LLMs)}} \\
    Gemini 3.1 Pro \cite{gemini31pro} & -- & 0.261 & 0.739 & $-$0.478 & 0.737 & 0.707 & 0.322 & 0.385 & 0.535 & 0.612 & 0.412 & 0.200 & 0.554 \\
    \midrule
    \multicolumn{14}{l}{\textit{Same checkpoints, refined prompt wording}} \\
    Pengi$^{\dagger}$ & 323M & 0.194 & 0.014 & 0.179 & 0.014 & 0.330 & 0.135 & 0.195 & 0.146 & 0.301 & 0.109 & 0.192 & 0.134 \\
    Qwen-Audio-Chat & 8.4B & 0.509 & 0.491 & 0.018 & 0.498 & 0.377 & 0.495 & $-$0.119 & 0.608 & 0.405 & 0.494 & $-$0.089 & 0.597 \\
    Qwen2-Audio-Instruct & 8.4B & 0.000 & 1.000 & $-$1.000 & \textbf{1.000} & 0.047 & 0.923 & $-$0.876 & 0.383 & 0.040 & 0.936 & $-$0.896 & 0.417 \\
    \bottomrule
  \end{tabular}
  }
  }
\end{table*}

%% file: section/6-discussion.tex
\section{Discussion}
\label{sec:discussion}

Our results highlight both the promise and limitations of current LALMs when evaluated through auditory illusions. The divergences between human and model perception vary systematically by illusion type. For physics-based cases, only a few models (e.g., MuLLaMa, Audio Flamingo 3) reproduced human-like misperceptions, while most—including recent systems such as Gemini 3.1 Pro and Kimi-Audio-Instruct—aligned with the physical signal. Illusions that required the integration of acoustic and prior-based cues elicited the most human-like responses, but the alignment remained partial (at best an ISI of 0.505) and, for several models, reversed the sign observed on physics-based illusions, suggesting that it is driven by linguistic and musical priors rather than by shared low-level auditory processing. The same checkpoint could also shift from moderate to near-complete signal fidelity under a reworded prompt. Notably, many models showed unstable performance even under control conditions, indicating that LALMs are still at an early stage of development.

These findings have broader implications for cognitive science and machine learning. From a cognitive perspective, illusions are a powerful diagnostic tool for probing auditory mechanisms, revealing which aspects of perception models capture and which remain elusive. 
From an engineering perspective, illusion susceptibility offers a complementary evaluation axis. It is important to note that the primary goal of \texttt{AIB} is to measure cognitive alignment rather than general task performance. The desirability of alignment depends on the application: In Human-Computer Interaction, higher alignment (high ISI) may support more natural and empathetic interaction. Conversely, in safety-critical settings or precision measurement tasks, reduced susceptibility (low ISI) may be preferable. Thus, \texttt{AIB} serves as a diagnostic tool to position models along this spectrum. We offer this benchmark as an initial step that may support the development of both cognitively grounded and task-appropriate LALMs.

%% file: section/7-conclusion.tex
\vspace{2mm}
\paragraph{Conclusion}
\label{sec:conclusion}
We introduced an auditory illusion benchmark for LALMs, pairing controlled human studies with systematic model evaluation. While models capture low-level regularities, they diverge in knowledge-driven illusions, underscoring limits in perceptual alignment. By framing illusions as diagnostic tools, we offer a complementary perspective for evaluating audio models and point to future work in expanding coverage, strengthening human baselines, and exploring reasoning-oriented training.  
We also highlight the value of illusions as probes of human-likeness in machine perception. Similar to visual benchmarks, auditory illusions reveal whether models reflect human biases or adhere strictly to the signal. We hope this benchmark can serve as a useful step toward advancing LALMs beyond recognition accuracy to more cognitively aligned auditory understanding.